\documentclass[twocolumn,showpacs,preprintnumbers,amsmath,amssymb]{revtex4}
\usepackage{tabularx,graphicx}

\usepackage{color}
\usepackage{hyperref}
\hypersetup{
    colorlinks=true,
    linkcolor=blue,
    filecolor=blue,      
    urlcolor=blue,
}

\begin{document}
%\documentstyle[aps]{revtex}
%\documentstyle[preprint,aps]{revtex}
%\begin{document}

\newcommand{\beq}{\begin{equation}}
\newcommand{\eeq}{\end{equation}}
\newcommand{\beqn}{\begin{eqnarray}}
\newcommand{\eeqn}{\end{eqnarray}}
\newcommand{\bmath}{\begin{subequations}}
\newcommand{\emath}{\end{subequations}}
\newcommand{\bra}[1]{\langle #1|}
\newcommand{\ket}[1]{|#1\rangle}

%\draft
\title{The disappearing momentum of the supercurrent in the superconductor to normal phase
transformation}
\author{J. E. Hirsch }
\address{Department of Physics, University of California, San Diego,
La Jolla, CA 92093-0319}

\begin{abstract} 
A superconductor in a magnetic field has surface currents that prevent the magnetic field from penetrating
its interior. These currents carry kinetic energy and mechanical momentum. 
When the temperature is raised and  the system becomes normal  the currents disappear.
Where do the kinetic energy and mechanical momentum of the currents go, and how? 
Here we propose that the answer to this question reveals a key necessary condition for
materials to be superconductors, that is not part of conventional BCS-London theory:
{\it superconducting materials need to have hole carriers}.
 \end{abstract}
\maketitle

 A superconductor in a magnetic field has shielding currents that keep magnetic field lines
 out of the superconductor except within a surface layer of thickness $\lambda_L$, the
 London penetration depth. It was discovered in 1914 by Kammerlingh Onnes \cite{onnes} that
 when the magnetic field exceeds a critical value $H_c$  that depends on temperature,
 the system becomes normal and the shielding currents disappear.  In this paper we discuss what happens to the kinetic energy and 
 mechanical momentum of the shielding currents when the system becomes normal, and how it happens, and argue
  that it has fundamental unrecognized implications
 for the understanding of superconductivity. We discuss only type I superconductors.
 
 Until the discovery of the Meissner effect in 1933 \cite{meissner} it was generally believed that 
 superconductors were nothing more than `perfect conductors' with zero resistivity.
 Within this point of view, when superconductors in the presence of a magnetic field became normal by 
  raising the temperature above the critical temperature for the given applied field, or by raising the applied field above the
  critical field for the given temperature, the
 resistivity would become finite and the shielding currents would decay by the usual scattering 
 processes in normal metals, i.e. phonons and impurities. This would cause the kinetic energy
 of the shielding currents to be dissipated as Joule heat in an irreversible way, and the mechanical momentum of
 the shielding currents to be
 transferred to the body as a whole through the same scattering processes that dissipate the energy and bring the current to 
 a halt. Within this point of view, if subsequently the system would be cooled again 
  it was expected that the 
 shielding currents would not reappear, rather that the magnetic field would remain in the interior of the body
 as it became superconducting again.
 
Meissner and Ochsenfeld's 1933 discovery \cite{meissner} however showed than on lowering the temperature the shielding currents are restored and the
magnetic field is expelled. This suggested (but did not prove) that the kinetic energy of the shielding currents
was in fact not lost to Joule heat as the system became normal, but rather became stored somewhere where
it could be subsequently retrieved and used to propel the shielding currents when the system was cooled again.
Indeed very precise experiments by Keesom and coworkers \cite{keesom} showed that in the process of the
system becoming normal and the shielding currents decaying to zero $no$ irreversible Joule heating occurs.
The kinetic energy of the supercurrents is used up in paying for the difference in free energies between normal
and superconducting states, as first discussed by H. London \cite{londonh}.

A conundrum that didn't exist before was thus created by Meissner's discovery: if there are no collision processes that dissipate Joule heat in the
superconductor-to-normal transition in the presence of a magnetic field, what happens
to the mechanical momentum of the disappearing current? The kinetic energy of the current is `stored' in the normal state electronic
state, but
its momentum is not. Of course the
only possible answer is that the momentum of the current is transmitted to the body as a whole.
But what is the physical mechanism by which this transfer of momentum happens with no energy transfer and no energy
dissipation? Surprisingly this basic and fundamental question has $never$ been
asked (nor answered) in the extensive literature on superconductivity since 1933 (213,616 papers according
to the Web of Science).

How do we actually know that the shielding supercurrents carry mechanical momentum? Because
it is expected theoretically and has been verified experimentally  by measuring the
gyromagnetic effect in superconductors \cite{gyro}: upon applying a magnetic field to a spherical or cylindrical
superconductor hanging from a thread, shielding currents develop
$and$ the body as a whole starts to rotate to keep the total angular momentum zero. The measured angular momentum of the body as a whole 
corresponds precisely to what is expected if the mechanical momentum density of the shielding current
$\vec{\mathcal{P}}$ is given by
\beq
\vec{\mathcal{P}}=\frac{m_e}{e}\vec{J}
\eeq
where $\vec{J}$ is the current density, $m_e$ the bare electron mass and $e$ ($<0$) the electron charge.
For applied magnetic field $H$, $J=c/(4\pi\lambda_L)H$, with $\lambda_L$ the London penetration depth.

          \begin{figure}
 \resizebox{8.5cm}{!}{\includegraphics[width=6cm]{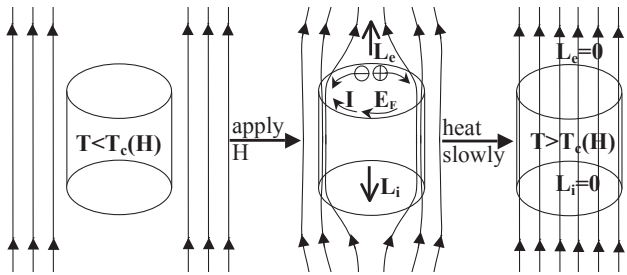}}
 \caption { When a magnetic field is applied to the superconductor, a Faraday field $E_F$ is generated that prevents penetration of the
 magnetic field beyond a London penetration depth of the surface of the superconductor. $E_F$ is clockwise as seen from the top in the
 Figure. $E_F$ drives electrons in the supercurrent to flow counterclockwise giving rise to a clockwise current that   generates a magnetic field opposite to the applied one, and
 $E_F$ also drives the positive ions in the body to move clockwise causing the body to rotate rigidly clockwise. The total angular momentum of electrons plus 
 ions $\vec{L}_e+\vec{L}_i$ is zero at all times. When the temperature is raised above $T_c(H)$, the system becomes normal, the magnetic field penetrates the interior,
 the supercurrent stops flowing $and$ the body stops rotating. The Faraday field generated in this latter process points in the same clockwise
 direction as the one generated when the field is applied.}
 \label{figure1}
 \end{figure}

The total momentum of the shielding currents will be zero, but the angular momentum will not. It  is  given by
\beq
\vec{L}_e= \int d^3r \vec{r}\times \vec{\mathcal{P}}(\vec{r})
\eeq
and an opposite angular momentum has to be acquired by the body as a whole. If we envision a process where the superconductor is
initially at rest  without magnetic field, application of a magnetic field will both induce the shielding
currents with their angular momentum $\vec{L}_e$  and impart opposite
angular momentum to the body as a whole $\vec{L}_i=-\vec{L}_e$. Both processes can be simply
understood as arising from the force imparted by the Faraday electric field induced as the
magnetic field is applied, counterclockwise for the negative electrons in the shielding currents and clockwise for the positive ions in the body as seen from the direction where the magnetic field points,
as shown schematically in Fig. 1.
The total angular momentum of the system (electrons plus ions) remains zero if the system is 
charge neutral.

As we subsequently slowly raise the temperature and the system becomes normal, the shielding currents stop and the rotation of the body has to
stop also,  so that the total angular momentum remains zero, now with $\vec{L}_e=\vec{L}_i=0$. This $cannot$ be understood as arising from force imparted by the Faraday electric field generated as the magnetic
field lines penetrate the body.
Quite the contrary, the Faraday electric  field acts in the same clockwise direction direction as when the magnetic field was first applied,
trying to   restore both the rotation of the body and
the flow of shielding currents. How then does the body stop rotating?

There is no microscopic theory that describes the process of the superconductor-normal transition
in the presence of a magnetic field (nor the reverse transition) within the conventional theory
of superconductivity. These problems have been studied   using the
phenomenological time-dependent Ginzburg-Landau formalism \cite{tdgl1,tdgl2,dorsey,goldenfeld}. 
Within this formalism Eilenberger has shown \cite{eilen} that when the superfluid electron density decreases
its mechanical momentum is transferred to the normal electrons, and according to Eilenberger
``this momentum then decays with the transport
relaxation time $\tau$''. Clearly this cannot be correct since it would lead to irreversible Joule heating
which is not observed \cite{keesom}. {\it How then do electrons in the supercurrent
 transfer their mechanical momentum to the ionic lattice
without energy dissipation?}

Consider Bloch electrons in the weak binding approximation moving in a perfect crystal.
Electrons interact with the crystal potential through its Fourier components $U_{\vec{K}}$
where $\vec{K}$ are reciprocal lattice vector. Electrons near the bottom of the band
are only weakly affected by the lattice potential, since the energy of an electron
scattered from $\vec{k}$ to $\vec{k}\pm \vec{K}$ will be vastly higher than 
$\epsilon_k^0$, the free electron energy for $\vec{k}$ near the bottom of the band. Instead, electrons near the top of the band
are strongly affected by the lattice potential since the energies $\epsilon_k^0$ and 
$\epsilon_{k \pm K}^0$ will be nearly equal for some reciprocal lattice vector(s)  $\vec{K}$.
In affecting the electronic state, the lattice transfers momentum to the electron. By Newton's 
third law, the electron transfers momentum to the lattice. This indicates that the electrons
that are most effective in tranferring mechanical momentum to the lattice without energy dissipation
are {\it electrons near the top of the band}.       Since superconducting electrons becoming normal in the presence of a magnetic field need to transfer
mechanical momentum to the lattice without energy dissipation, we conclude that
{\it  materials that can become supeconductors need to have electrons near the top of electronic energy bands}.
In other words, $holes$.

More generally, consider the semiclassical equation of motion for an electron of wavevector $\vec{k}$:
\beq
\hbar \frac{d\vec{k}}{dt}=\vec{F}_{ext}
\eeq
where $\vec{F}_{ext}$ is an  external force. The $total$ force acting on the electron is
\beq
m_e \frac{d \vec{v}_{\vec{k}}}{dt}=\vec{F}_{ext}+\vec{F}_{L}
\eeq
with
\beq
\vec{v}_{\vec{k}}=\frac{1}{\hbar}\frac{\partial \epsilon_{\vec{k}}}{\partial\vec{k}}
\eeq
and $\vec{F}_{L}$ the force that the lattice exerts on the electron, given by
\beq
\vec{F}_{L}=(m_e \frac{1}{\hbar^2}     \frac{\partial^2 \epsilon_{\vec{k}}}{\partial\vec{k}\partial\vec{k}}   -1)\vec{F}_{ext}  .
\eeq
By Newton's third law, the electron in turn exerts a force on the lattice
\beq
\vec{F}_{on-L}=-\vec{F}_{L}=(1-m_e \frac{1}{\hbar^2}     \frac{\partial^2 \epsilon_{\vec{k}}}{\partial\vec{k}\partial\vec{k}} )\vec{F}_{ext}  .
\eeq
which transfers momentum from the electron to the lattice. The largest momentum transfer will occur when the second derivative term
in Eq. (7) is negative, which happens when electrons are near the top of a band, i.e. when there is $hole$ conduction.

          \begin{figure}
 \resizebox{8.5cm}{!}{\includegraphics[width=6cm]{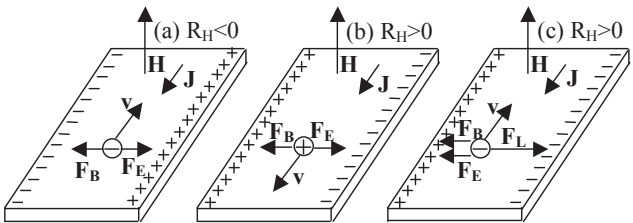}}
 \caption {(a) Hall effect when carriers are electron-like. Electric and magnetic forces
$F_E$ and $F_B$   on the charge carriers in direction perpendicular to the current $\vec{J}$ point in opposite directions and cancel each other. (b) Hall effect when
 carriers are hole-like. Electric and magnetic force on $positive$ charge
 carriers (holes)  cancel
 each other. (c) Reinterpretation of the forces for case (b): since the mobile
 charge carriers are always electrons, electric and magnetic forces point in the
 same direction and need to be cancelled by a the force $F_L$ exerted
 by the lattice on electrons. }
 \label{figure1}
 \end{figure}

Consider next the motion of electrons in crossed electric and magnetic fields, as shown in Fig. 2.
The Hall coefficient is defined as $R_H=E_y/(J_xH)$, with $J_x$ the current in the longitudinal direction, $E_y$ the electric field in the transverse direction and $H$ the magnetic field in the perpendicular direction. By setting
$J_x=nev$, with n the concentration of carriers of charge e moving with drift velocity v, electric and magnetic Lorentz forces $F_E$ and $F_B$  are balanced for
$E_y=(v/c)H$ ($F_E=eE_y=F_B=(ev/c)H$) and it follows that the Hall coefficient is given by
\beq
R_H=\frac{1}{nec}
\eeq
with the sign as shown in Fig. 2(a) assuming the mobile carriers are electrons. 
As shown  by Ashcroft and Mermin \cite{am}, Eq. (8) holds for Bloch electrons, with the current and number
of carriers given by
\bmath
\beq
\vec{J}=\int_{occ} \frac{d^3k}{4\pi^3} \frac{1}{\hbar}\frac{\partial \epsilon_{\vec{k}}}{\partial \vec{k}}
\eeq

\beq
n=\int_{occ} \frac{d^3k}{4\pi^3}
\eeq
\emath
if all occupied k-space orbits are closed, which occurs when the band is closed to empty. In this case, electric
and magnetic forces on electrons are balanced on average as shown in Fig. 2(a) and  no net force is exerted by the lattice on electrons nor by electrons on the lattice
as the current flows.

On the other hand, if all $unoccupied$ k-space orbits are closed, which occurs when the
band is almost full, the Hall voltage has opposite sign and the Hall coefficient is given by \cite{am}
\beq
R_H=\frac{1}{n_h|e|c}
\eeq
with
\bmath
\beq
\vec{J}=\int_{unocc} \frac{d^3k}{4\pi^3} \frac{1}{\hbar}\frac{\partial \epsilon_{\vec{k}}}{\partial \vec{k}}
\eeq
\beq
n_h=\int_{unocc} \frac{d^3k}{4\pi^3}
\eeq
\emath
According to Fig. 2(b), the electric and magnetic forces $on$ $holes$ are equal and opposite and no net force
results. However, this is misleading, since electric and magnetic forces act on electrons and not on holes.
As shown in Fig. 2(c), electric and magnetic forces on $electrons$ point in the same direction,
and a lattice counterforce $F_L=F_E+F_B$  is exerted by the lattice on the electron to keep its trajectory along the direction
of the current.
This in turn implies that when $R_H>0$ {\it a steady force $-F_L=-(F_E+F_B)$ is exerted by the electron on the
lattice as the current flows.}

This force exerted by the carriers on the lattice when current flows transfers momentum
from the carriers to the lattice without energy dissipation. We are not aware of any
other physical mechanism by which charge carriers in a solid can transfer momentum to the lattice without
scattering processes that lead to energy dissipation. Consequently, we propose
that the process shown in Fig. 2(c) describes the essential physics of how
the momentum of the supercurrents is transferred to the body as a whole
when a superconductor in a magnetic field makes a reversible transition 
to the normal state.

A detailed realization of this mechanism is provided by the theory of
hole superconductivity \cite{holesc}. Within that theory, superconducting electrons
reside in mesoscopic orbits of radius $2\lambda_L$ \cite{bohr}, with $\lambda_L$
the London penetration dept. When carriers go from normal to superconducting,
their orbits expand from radius $k_F^{-1}$ to radius $2\lambda_L$ driven by lowering of quantum kinetic energy, and if
a magnetic field is present they acquire through the magnetic Lorentz force the
angular velocity required to provide a dynamical explanation of the origin of the
Meissner current \cite{sm}. We now explain how the transfer of momentum to the lattice
occurs for the case of interest here, for a planar geometry for simplicity.

Figure 3 shows schematically the large orbits in the superconducting state (large overlapping circles) centered below 
the phase boundary line (horizontal dotted line), and the small orbits (small nonoverlapping circles) in the normal state above the
phase boundary line, in the presence of the critical magnetic field $H_c$ in the normal region pointing out of the
paper. As the phase boundary moves down into the superconducting region, large orbits right at the phase boundary shrink, as shown
by the circles of diminishing radius. This causes negative charge to be transferred out of a boundary layer
of thickness $\lambda_L$ above the phase boundary, and gives rise to a $backflow$ of electrons moving
in the positive $\hat{x}$ direction \cite{dynann}, indicated by the vertical arrows labeled `electron backflow'. This corresponds 
equivalently to a current flowing in the $-\hat{x}$ direction.

          \begin{figure}
 \resizebox{8.5cm}{!}{\includegraphics[width=6cm]{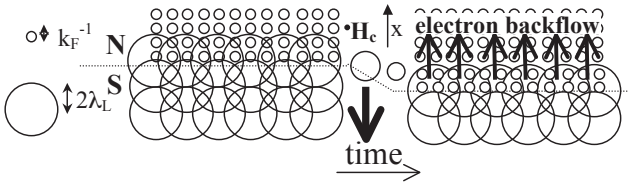}}
 \caption {Normal-superconductor phase boundary advancing in the $-\hat{x}$
 direction. Large orbits of superconducting electrons near the phase boundary
 extend into the normal region. As these orbits shrink, it causes a backflow of normal
 electrons flowing in the  $+\hat{x}$  direction as indicated by the vertical arrows. 
 Magnetic field $H_c$ points out of the paper.}
 \label{figure1}
 \end{figure} 

           \begin{figure}
 \resizebox{8.5cm}{!}{\includegraphics[width=6cm]{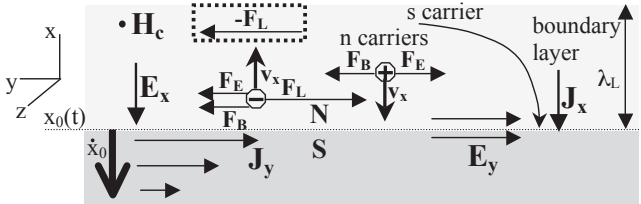}}
 \caption {Schematics of currents and fields for the situation depicted in Fig. 3 where
 the normal phase is advancing in the $-\hat{x}$ direction.
 The Faraday field $E_y$ points in the negative $\hat{y}$ direction. 
 The normal electron backflow depicted in the positive $\hat{x}$ direction
 corresponds to a normal current $J_x$ flowing in the $-\hat{x}$ direction
 within a boundary layer of thickness $\lambda_L$ from the phase boundary.
 If the normal carriers (n carriers) are hole-like, the lattice exerts force $F_L$
 on the carriers in the $\hat{y}$ direction and correspondingly the carriers
 exert a force $-F_L$ on the lattice (shown in the dotted rectangle) in the $+\hat{x}$ direction.
 The momentum in the $\hat{y}$ direction imparted by the backflow on the lattice equals the momentum in the $\hat{y}$ direction of the carriers of the supercurrent $J_y$
 becoming normal
 that is lost through the action of the magnetic Lorentz force as the orbits shrink.
}
 \label{figure1}
 \end{figure} 
 
Figure 4 shows all the currents and fields schematically, in a situation where the phase boundary
located at $x_0(t)$  is moving down into the
superconducting phase at a uniform speed $\dot{x}_0$. A Faraday electric field $E_y$ pointing in the $-\hat{y}$ direction is generated at and in the
neighborhood of the phase boundary due to the changing magnetic flux, given by
\beq
E_y=\frac{\dot{x}_0}{c}H_c
\eeq
The backflow current $J_x$ flowing in the $-\hat{x}$ direction is assumed to be hole-like, corresponding to the situation in Figure 2(c), and has magnitude $J_x=n_h|e|\dot{x_0}$, witn $n_h$ the hole carrier concentration.
The forces on electrons are balanced by a force $F_L$ exerted by the lattice on the electrons, and the electrons exert a counterforce on the 
lattice $-F_L=-(F_E+F_B)$ in the $+\hat{y}$ direction, shown in Fig. 4 in the dotted rectangle. In addition, the Faraday field exerts a force
$F_E$ on the lattice in the negative $\hat{y}$ direction (not shown in Fig. 4).   The net force on the lattice
per carrier is then
$-F_E=eE_y$, pointing in the $+\hat{y}$ direction. 

 In more detail the balance is as follows. The superconducting electrons at the boundary have velocity in the
$+\hat{y}$ direction   \cite{dynann}
\beq
\vec{v}_y=      \frac{e}{m_ec}\  \lambda_L   H_c \hat{y}
\eeq
and kinetic energy
\beq
\epsilon_{kin}=\frac{1}{2}m_e v_y^2=\frac{e^2 \lambda_L^2}{2m_e c^2}H_c^2=\frac{1}{n_s}\frac{H_c ^2}{8\pi}
\eeq
using that $1/\lambda_L^2=4\pi n_s e^2 / m_e c^2$. 
An electron going from superconducting to normal shrinking its orbit
effectively moves
at a high speed $v_x$ in the negative $\hat{x}$ direction a distance $\lambda_L$ in time $\lambda_L/v_x$ under
the action of the magnetic Lorentz force $(e/c)v_xH_c$, thereby changing its momentum by
\beq
\Delta p_y=\frac{e}{c} \lambda_L H_c .
\eeq
This change in momentum is in the $-\hat{y}$ direction, and exactly cancels the momentum in the 
$+\hat{y}$ direction that  the electron had initially
being  part of the Meissner current.  
This assumes that the speed $v_x$ is much larger than $\dot{x}_0$, so that
the effect of $E_y$ over this short time (which applies a force in the opposite ($+\hat{y}$) direction) can be neglected.
This then explains how the Meissner current comes to a halt without
dissipation. The kinetic energy that the electron lost Eq. (14) is the condensation energy per electron, i.e.what it
costs to bring the electron from the superconducting to the normal state. 
Multiplying Eq. (14) by the number of superfluid carriers per unit volume $n_s$ yields the condensation
energy per unit volume $H_c^2/(8\pi)$ \cite{londonh}.

 At the same time the `backflow' normal electrons move at speed $\dot{x}_0$ in the $+\hat{x}$ direction,
 and traverse the boundary layer distance $\lambda_L$ in time $\Delta t=\lambda_L/\dot{x}_0$. 
 The $net$ momentum imparted to the lattice in this process is $\Delta p_y=F_E\Delta t=eE_y\Delta t=(e/c)\lambda_L H_c$ in the $+\hat{y}$ direction, the
 same momentum that a superconducting electron becoming normal lost,  Eq. (15). In this way the momentum of the supercurrent is transmitted to the lattice without dissipation. 
 For the cylindrical body, the end result of this process when the entire system has become normal
is that there is no more supercurrent flow and no rotation of the body.

Note that the backflow  current $J_x$  is in direction exactly perpendicular to the phase boundary because the forces
in the $\hat{y}$ direction are balanced if the normal state carriers are holes. Because the backflow
occurs only over a small boundary layer
of thickness $\lambda_L$  it 
will give rise to no energy dissipation assuming the mean free path is larger than $\lambda_L$. Instead,
if the normal state carriers were electrons rather than holes the situation would be as depicted in Fig. 5.
There would be no force by the lattice on the electrons and the backflowing  electrons would acquire a tangential
velocity in the $+\hat{y}$ direction, and this current would die out by scattering transmitting momentum to
the lattice and dissipating Joule heat. Thus within this scenario the observation that no Joule heat is dissipated
is only compatible with the normal state carriers being hole-like.

            \begin{figure}
 \resizebox{8.5cm}{!}{\includegraphics[width=6cm]{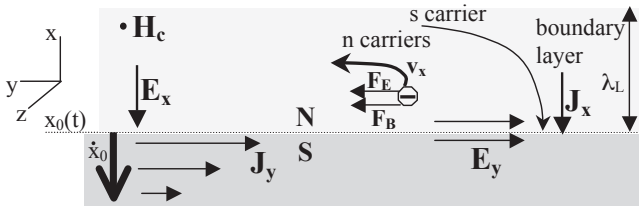}}
 \caption {Same as Fig. 4 if the normal state carriers are electrons. There is no force by the lattice on the backflow electrons
hence the force in the $\hat{y}$ direction is not balanced. A tangential flow of normal electrons would take place,
and momentum would be transferred to the lattice through scattering processes leading to energy dissipation.
}
 \label{figure1}
 \end{figure}

 In summary, we have pointed out in this paper that the fact that the superconductor-normal transition is found experimentally to be reversible,
 hence occurs without Joule heat dissipation, poses a conundrum that is not addressed in the conventional theory of superconductivity: how
 does the momentum of the supercurrent get transferred to the superconducting body without energy dissipation? We have pointed out
 that Bloch's  theory of electrons in crystals shows that electrons near the top of the band are effective in transmitting their momentum to the
 lattice in a reversible way  because they undergo Bragg scattering, and electrons near the bottom of the band are not.
 For electrons moving in crossed electric and magnetic fields we have pointed out that when the Hall coefficient is negative no net momentum
 transfer between electrons and the lattice occurs, while if the Hall coefficient is positive a net momentum transfer between electrons and the lattice
 necessarily occurs. Finally, we have proposed a specific scenario using physical elements from the theory of hole superconductivity that
 explains how the Meissner current stops and the momentum is transferred to the body as a whole without energy dissipation
 when the superconductor goes normal.

 From its inception \cite{holesc1} the theory of hole superconductivity proposed that hole carriers are indispensable for superconductivity to occur.
 Over the years we have discussed many different reasons in favor of this hypothesis \cite{holesc}. The additional reason discussed in this paper
is arguably the most compelling one.
\acknowledgements
The author is grateful to Nigel Goldenfeld for calling Ref. 10 to his attention.
   
\end{document}